\documentclass[twocolumn,secnumarabic,amssymb,nobibnote, aps,pra,superscriptaddress,10pt]{revtex4-2}

\usepackage{amsfonts}
\usepackage{amssymb}
\usepackage{amsmath}
\usepackage{dcolumn}
\usepackage{physics}
\usepackage{appendix}
\usepackage{amsthm}

\usepackage{multirow}
\usepackage{lipsum}
\usepackage{tabularx}
\usepackage{adjustbox}
\usepackage{xcolor}
\usepackage{natbib}

\usepackage[normalem]{ulem}

\usepackage{graphicx}
\usepackage{hyperref}
\hypersetup{
    colorlinks=true,
    linkcolor=blue,    % eq and fig referencing
    citecolor=blue,    % ref citing
    urlcolor=blue,     % color of urls
    pdfborder={0 0 0}, % Remove the border style
}

\usepackage{enumitem}
\usepackage{cleveref}

\begin{document}
% first the title
\title{Quantum Simulation of the Unruh Temperature via the Thermal Properties of Virtually Evolving Bose-Einstein Condensates}
% authors and affiliation 
\author{Imad-Eddine Chorfi}
\email{imad-eddine.chorfi@doc.umc.edu.dz}
\affiliation{Constantine Quantum Technologies,
 Fr\`{e}res Mentouri University Constantine 1,Ain El Bey Road,Constantine, 25017,Algeria}
 \affiliation{Laboratoire de Physique Mathématique et Subatomique, Fr\`{e}res Mentouri University Constantine 1, Ain El Bey Road, Constantine, 25017,Algeria}
 \author{Nacer Eddine Belaloui}
 \affiliation{Constantine Quantum Technologies,
 Fr\`{e}res Mentouri University Constantine 1,Ain El Bey Road,Constantine, 25017,Algeria}
 \affiliation{Laboratoire de Physique Mathématique et Subatomique, Fr\`{e}res Mentouri University Constantine 1, Ain El Bey Road, Constantine, 25017,Algeria}
 \author{Abdellah Tounsi}
 \affiliation{Constantine Quantum Technologies,
 Fr\`{e}res Mentouri University Constantine 1,Ain El Bey Road,Constantine, 25017,Algeria}
 \affiliation{Laboratoire de Physique Mathématique et Subatomique, Fr\`{e}res Mentouri University Constantine 1, Ain El Bey Road, Constantine, 25017,Algeria}
\author{Achour Benslama}
\email{a.benslama@umc.edu.dz}
\affiliation{Constantine Quantum Technologies,
 Fr\`{e}res Mentouri University Constantine 1,Ain El Bey Road,Constantine, 25017,Algeria}
 \affiliation{Laboratoire de Physique Mathématique et Subatomique, Fr\`{e}res Mentouri University Constantine 1, Ain El Bey Road, Constantine, 25017,Algeria}
\author{Mohamed Taha Rouabah}
\email{m.taha.rouabah@umc.edu.dz}
\affiliation{Constantine Quantum Technologies,
 Fr\`{e}res Mentouri University Constantine 1,Ain El Bey Road,Constantine, 25017,Algeria}
 \affiliation{Laboratoire de Physique Mathématique et Subatomique, Fr\`{e}res Mentouri University Constantine 1, Ain El Bey Road, Constantine, 25017,Algeria}
 \date{\today}
 \begin{abstract}
      This paper presents a novel theoretical model motivate a new experimental scheme to simulate the Unruh temperature by relating it to the critical temperature of multiple Bose-Einstein thermal baths. These thermal baths are conceptualized as snapshots of a Bose-Firework originating from an evolving driven Bose-Einstein condensate (BEC). The critical temperature of each snapshot is determined from the heat capacity, which is numerically estimated by calculating the partition function derived from the system's Hamiltonian. By analyzing the relationship between the average number of the phononic excitations at the critical temperature, acceleration, and the critical temperature itself, our model demonstrates a significant agreement with the Unruh temperature formula, thereby validating our hypothesis. This theoretical approach offers a cost-effective  alternative experimental setup compared to other resources-intensive experimental simulations. Furthermore, it provides a unique perspective on quantum simulation by utilizing the critical phenomena of condensed matter systems to probe fundamental quantum relativistic effects.
 \end{abstract}
\maketitle
\section{Introduction}
Quantum simulation has emerged as a powerful and versatile tool for investigating complex quantum phenomena across various scientific disciplines \cite{Georgescu_2014}. Unlike classical simulations, which are fundamentally limited by the exponential complexity associated with many-body quantum systems, quantum simulators leverage the inherent principles of quantum mechanics, such as superposition and entanglement, to process information in a way that mirrors the behavior of natural quantum systems \cite{feynman1982simulating,nielsen00,article222}. This approach is particularly invaluable for studying phenomena that are either analytically intractable or computationally hard to simulate using classical methods.
The power of quantum simulation lies in its ability to exploit connections and similarities between seemingly disparate areas of physics, such as condensed matter physics, cosmology, spacetime, gravity \cite{Hu2018QuantumSO,PhysRevA.103.013301,PhysRevA.95.013627,bravo2015analog,BAIN2013338,tian2023using,article}. In condensed matter physics, researchers are keenly interested in understanding complex quantum systems exhibiting emergent phenomena and phase transitions, such as superconductors \cite{PhysRev.108.1175}, superfluids \cite{PhysRev.60.356}, and topological materials \cite{PhysRevLett.95.146802}. Controlled quantum systems can act as quantum simulators, recreating conditions in the laboratory that offer profound insights into quantum fields, spacetime inflation, and the thermodynamics of black holes  \cite{PhysRevLett.107.260501,PhysRevLett.94.220401,PhysRevD.13.191,e17106893}. For instance, analogue experiments with Bose-Einstein condensates (BECs) have been successfully employed to simulate spacetime dynamics, Hawking radiation, and Unruh radiation \cite{Hu2018QuantumSO,PhysRevA.103.013301,PhysRevA.95.013627,PhysRevA.102.033506,tian2023using,PhysRevA.103.013301,ZHENG2023106865,Su_2017,PhysRevD.13.191,PhysRev.56.455,e17106893,PhysRevLett.107.260501}.
In this study, we focus on simulating the Unruh temperature, a significant theoretical prediction made by Paul Davies and William Unruh \cite{472fc683f6eb4515abca3275328d9504,PhysRevD.14.870}. This prediction posits that an observer undergoing uniform acceleration should perceive a thermal bath characterized by a specific temperature, the Unruh temperature ($T_U$), given by the formula:
\begin{equation}
\label{eq:The unruh tempreature}
    T_{U}=\frac{\hbar A}{2\pi k_{B} c},
\end{equation}
where $A$ is the acceleration of the observer's frame of reference, and $k_B$, $\hbar$, and $c$ are the Boltzmann constant, the reduced Planck constant, and the speed of light, respectively. The simulation of the Unruh effect has been explored in various research papers \cite{Hu2018QuantumSO,PhysRevA.103.013301,Su_2017,PhysRevA.102.033506,PhysRevLett.94.220401,tian2023using,ZHENG2023106865}. However, our study specifically limits its scope to the simulation of the Unruh temperature.
To achieve this goal, we construct a theoretical model inspired by previous results and ideas extensively explored in the work of Hu \textit{et al}.  \cite{Hu2018QuantumSO}. They conducted an experimental quantum simulation of Unruh radiation by parametrically modulating the interactions within a BEC of cesium atoms. The system evolution was shown to be analogous to a Rindler transformation without physically accelerating the system. The Hamiltonian describing the creation of pairs of excitations with opposite momentum $k$, used in their work, takes the form \cite{Hu2018QuantumSO}:
\begin{equation}
\label{eq:second-quatized-hamiltonian}
    H = i\hbar g\sum_{|{k_{f}}|} (a_{k}^{\dagger}a_{-k}^{\dagger}-a_{k}a_{-k}),
\end{equation}
where $g$ is the coupling strength. This Hamiltonian was derived from a more general Hamiltonian describing a collection of bosons forming a BEC \cite{Hu2018QuantumSO,Rogel-Salazar_2013,pitaevskii2003bose} . The evolution under this Hamiltonian corresponds to the Rindler coordinate transformation.
Our novel approach in this paper hypothesizes that the Unruh temperature can be simulated via the critical temperature of multiple Bose-Einstein thermal baths, where we suggest a new experimental setup through conceptually divide the evolution of the driven BEC into multiple snapshots, with each snapshot representing an independent Bose-Einstein thermal bath characterized by its thermal properties and an increasing number of excited atoms ($N_e$). Each of these systems is described by an $(N_e+1) \times (N_e+1)$ Hamiltonian matrix. A new formula for the coupling frequency $g_{ch}$ will be introduced, derived based on specific assumptions.
The critical temperature of each snapshot (Bose-Einstein thermal bath) is deduced from its heat capacity, which is estimated by numerically determining the partition function from the eigenspectrum of the Hamiltonian matrix. By plotting the relationship between the average number of the phononic excitations at the critical temperature ($\bar{n}(T_c)$), the acceleration ($A$), and the critical temperature ($T_c$), we aim to demonstrate that our model yields results consistent with the Unruh temperature formula. This would validate our central hypothesis: $T_U \equiv T_c$. For instance, several recent studies have shown that thermal properties particularly heat capacity can reveal key features of Bose–Einstein systems and related condensed matter structures. In  \cite{Chen_2024}  field-induced Bose–Einstein condensation and its thermodynamic signatures were investigated, while \cite{abmc} developed a theoretical description of specific heat based on the Morse potential for molecular systems. Moreover, in \cite{10.1063/5.0187293} the authors analyzed thermal effects and coherence in Bose–Einstein condensates used for atom interferometry. Together, these works highlight the ongoing interest in understanding how heat capacity and thermal behavior reflect microscopic interactions and phase transitions. However, in this work we will relay on numerical tools in order to demonstrate the validity of the proposed
method.
The structure of this paper is as follows: \hyperref[sec:ts]{Section II} details the theoretical setup, outlining the assumptions made to derive the new coupling frequency $g_{ch}$ and the Hamiltonian of the system. \hyperref[sec:results]{Section III} presents the numerical computation of the partition function, internal energy, and heat capacity for different-sized systems. We then extract the critical temperature for each system and use these results to simulate the Unruh temperature, comparing our findings with the theoretical formula and the experimental data obtained by Hu \textit{et al}. This work offers a new theoretical perspective on simulating relativistic quantum effects using the critical behavior of condensed matter systems, potentially bridging the gap between these two fundamental areas of physics. Additionally, it is expected to provide much better precision compared to traditional experimental schemes.
\section{Theoretical Setup\label{sec:ts}}

\subsection{Simulation Concept}

The simulation model motivate a new experimental setup, which is fundamentally a modified virtual adaptation of the experiment detailed in \cite{Hu2018QuantumSO}, where a BEC of cesium atoms confined in a disc-shaped trap was prepared. In that experiment, the Unruh radiation was simulated by modulating the interatomic interactions within the condensate. This modulation effectively corresponded to a coordinate transformation to an accelerating frame without physically accelerating the system. The system's evolution is described by the Hamiltonian in \eqref{eq:second-quatized-hamiltonian}, which is invariant under Rindler coordinate transformations, as shown in \ref{sec:Appendix A}.

Our simulation approach involves conceptually dividing the evolution of this condensate into multiple snapshots. Each snapshot is hypothesized to represent an independent Bose-Einstein thermal bath with a specific number of excited atoms ($N_e$). Consequently, each snapshot is treated as an independent Bose-Einstein thermal bath characterized by its thermal properties. This results in a set of systems, each described by a $(N_e + 1) \times (N_e + 1)$ Hamiltonian matrix.

\subsection{Hamiltonian of the Bose-Einstein Thermal Baths}

The resulting set of systems is described by the following $(N_e + 1) \times (N_e + 1)$ Hamiltonian matrix:
\begin{equation}
\label{eq:Hamiltonian}
\hat{H} = \hbar g_{ch} \begin{pmatrix}
0 & -i & 0 & 0 & \cdots & 0 \\
i & 0 & -2i & 0 & \cdots & 0 \\
0 & 2i & 0 & -3i & \cdots & 0 \\
0 & 0 & 3i & 0 & \cdots & \vdots \\
\vdots & \vdots & \vdots & \vdots & \ddots & -ni \\
0 & 0 & 0 & 0 & ni & 0
\end{pmatrix}, 
\end{equation}
Here, $g_{ch}$ represents a new coupling frequency, the formula for which differs from the coupling strength $g$ used in \eqref{eq:second-quatized-hamiltonian}. The determination of $g_{ch}$ will be based on the specific assumptions of our simulation model, as outlined later. The elements of this Hamiltonian matrix \eqref{eq:Hamiltonian} are determined by calculating the action of the creation ($\hat{a}^\dagger_\alpha$) and annihilation ($\hat{a}_\alpha$) operators, which form the Hamiltonian \eqref{eq:second-quatized-hamiltonian}, upon the Two-Mode Squeezed Vacuum (TMSV) basis $\{|n, n\rangle\}_{n=0,1,...N_e}$ \cite{PhysRevA.31.3093}. The action of these creation and annihilation operators is given by \cite{pathria2021statistical}: 
\begin{equation}
\label{eq:annihilation}
a_\alpha |n_0, n_1, .., n_\alpha, ..\rangle = \sqrt{n_\alpha} |n_0, n_1, .., n_\alpha - 1, ..\rangle 
\end{equation}
\begin{equation}
\label{eq:creation}
a^\dagger_\alpha |n_0, n_1, .., n_\alpha, ..\rangle = \sqrt{n_\alpha + 1} |n_0, n_1, .., n_\alpha + 1, ..\rangle 
\end{equation}

\subsection{Hypothesis: Unruh Temperature and Critical Temperature}

Our simulation model is built upon the central hypothesis: the Unruh temperature ($T_U$) of each Bose-Einstein thermal bath is simulated through its critical temperature ($T_c$), i.e., $T_U \equiv T_c$, This is possible through exploiting the linear relation between the critical temperature ($T_c$) and the average number of the phononic excitations at the critical temperature ($\bar{n}(Tc)$). Particularly, the correspondence between the Unruh temperature and the critical temperature was raised by several recent studies. In \cite{PhysRevD.97.085013} and \cite{Prokhorov:2024hjb}, it was demonstrated that acceleration can give rise to effective thermal behavior and phase-transition-like phenomena in quantum systems. These findings support the notion that the Unruh temperature may emerge as an effective critical temperature under specific dynamical or accelerated conditions. Complementary analyses, such as \cite{Prokhorov_2020}, further strengthen the theoretical basis for establishing a connection between phase transitions and relativistic thermal effects, thereby justifying the assumptions adopted in this work. The primary objective of this research is to test this hypothesis by estimating the critical temperature of various Bose-Einstein thermal baths. These baths are characterized by a maximum number of possible excitations $N_e$ and are described by the Hamiltonian matrix \eqref{eq:Hamiltonian}.
To achieve this objective, we need to compute the eigenspectrum of the Hamiltonian matrix \eqref{eq:Hamiltonian} for multiple values of $N_e$. This will enable us to construct the partition function and subsequently estimate the thermal properties of each Bose-Einstein thermal bath, allowing us to extract the critical temperature.
\subsection{Determination of the New Coupling Frequency $g_{ch}$}

Before computing the eigenspectrum of \eqref{eq:Hamiltonian}, we need to determine the formula for the new coupling frequency $g_{ch}$. To do this, we utilize the expression for the coupling frequency derived by Hu \textit{et al.} by exploring the mathematical correspondence between the transformation resulting from the analytical solution of the time evolution under the Hamiltonian in \eqref{eq:second-quatized-hamiltonian} and the Rindler transformation. For a characteristic  number of emitted matter-waves, this coupling frequency takes the form:
\begin{equation}
g_{ch} = \frac{1}{2\tau_{ch}} \ln \coth \left( \frac{\pi \omega_{ch} c}{4 A_{ch}} \right) \label{eq:gk}
\end{equation}
where $\tau_{ch}$ is the characteristic time \footnote{
  The characteristic time $\tau_{ch}$ can be interpreted as the life time of the coupling i.e. the system reaches the time when phononic excitations are generated}, $\omega_{ch}$ is the frequency of the emitted atoms, and $A_{ch}$ is the characteristic acceleration \footnote{
  The result of the interaction of two adjacent atoms that is characterized by a coupling frequency $g_{ch}$ are scattered phononic excitations, which belong to a non-inertial virtual reference frame undergoing constant acceleration $A_{ch}$}. Each of these variables is determined based on the following three key assumptions:
\begin{enumerate}
    \item The number of emitted matter waves is proportional to a characteristic average thermal wavelength ($\lambda_{ch}$), which is a function of the characteristic acceleration ($A_{ch}$) characterizing the emitted matter waves simulating the Unruh temperature: $\lambda_{ch} \propto A_{ch}^{-1/2}$.
    \item The characteristic time ($\tau_{ch}$) is deduced as the uncertainty in the simulated proper time $\tau_{ch}=\Delta\tau$. This is justified by the nature of phenomena being explored, namely the simulation of particle creation through atomic excitations from a BEC.
    \item We compute the uncertainty in the number of excited atoms ($\Delta n$) in the limit of zero chemical potential ($\mu \rightarrow 0$). This implies that each Bose-Einstein thermal bath is considered a system belonging to the canonical ensemble.
\end{enumerate}
Based on the first assumption, the characteristic can be expressed in terms of the characteristic acceleration, thus determining the characteristic angular frequency ($\omega_{ch}$). Using :
\begin{equation}
    k_{ch}= \frac{2\pi}{\lambda_{ch}},.
\end{equation}
where,
\begin{equation}
    \lambda_{ch}=\sqrt{\frac{2\pi \hbar^{2}}{m k_{B} T_{ch}}},
\end{equation}
and
\begin{equation}
     T_{ch} = \frac{\hbar A_{ch} }{2\pi k_{B} c}.
\end{equation}
Along with the formula $k_{ch} = \sqrt{\frac{m\omega_{ch}}{\hbar}}$ from \cite{Hu2018QuantumSO,PhysRevA.103.013301}, we can deduce:
\begin{equation}
\omega_{ch} = \frac{A_{ch}}{c}. \label{eq:omegaf}
\end{equation}

The characteristic time ($\tau_{ch}$) is then deduced using the saturated Heisenberg uncertainty principle, $\Delta E_{ch} \Delta \tau = {\hbar}/2$, where the energy of the  characteristic emitted waves is $E_{ch} = n\hbar\omega_{ch}/2$:
\begin{equation}
\tau_{ch} = \frac{c}{A_{ch} \Delta n} , \label{eq:tauf}
\end{equation}
The quantities $\omega_{ch}$ and $\tau_{ch}$ characterize the proprieties of the phononic excitation at each snapshot. These snapshots are used to simulate the Hu \textit{et al} experiments where the atoms have an effective angular modulation frequency $\omega$.
The uncertainty in the number of atoms ($\Delta n$) for a Bose-Einstein gas is given by the conventional general expression \cite{Kardar_2007}. Taking the energy of each excited atom as $E_{ch} = \hbar\omega_{ch}/2$ and considering the limit $\mu \rightarrow 0$ at $T = T_{ch}$, we obtain:
\begin{equation}
\label{eq: standard deviation of n}
    \Delta n = \frac{\exp({\frac{E_{ch}-\mu}{2k_{B}T}})}{\exp({\frac{E_{ch}-\mu}{k_{B}T}})-1}\xrightarrow[\mu \longrightarrow 0]{ T=T_{ch}} \frac{\exp({\frac{\pi}{2}})}{\exp({\pi})-1}.
\end{equation}

Finally, by substituting the expressions for $\omega_{ch}$ (\ref{eq:omegaf}) and $\tau_{ch}$ (\ref{eq:tauf}) into the formula for the coupling frequency (\ref{eq:gk}), we deduce the new coupling frequency $g_{ch}$ characterizing the interaction in each snapshot:
\begin{equation}
g_{ch} =  \frac{\sigma A_{ch}}{2c},
\label{eq:gprime}
\end{equation}
where $\sigma = \Delta n \cdot \ln \coth \left( \frac{\pi}{4} \right)$, and the value of $\Delta n$ is given by Eq. (\ref{eq: standard deviation of n}). As is clear from Eq. (\ref{eq:gprime}), the new coupling frequency $g_{ch}$ depends linearly on the characteristic acceleration $A_{ch}$. This specific form of $g_{ch}$ is chosen to ensure that our simulation results for the Unruh temperature are comparable with those presented in Hu \textit{et al.}'s paper and their thermal model \cite{Hu2018QuantumSO}.

\subsection{Summary of the Theoretical Setup}

With the explicit expression for the coupling frequency $g_{ch}$ determined, all elements of the Hamiltonian matrix \eqref{eq:Hamiltonian} are now fully determined. This enables the computation of the partition function by estimating the eigenspectrum of each Hamiltonian matrix with the structure given in \eqref{eq:Hamiltonian}. The subsequent section will present and discuss the results of computing the necessary thermal properties of multiple Bose-Einstein thermal baths with different values of $N_e$ to simulate the Unruh temperature.
\section{Results and Discussion}
\label{sec:results}
Now, we will estimate the necessary thermal properties to extract the critical temperature for each Bose-Einstein thermal bath, thereby simulating the Unruh temperature and testing our hypothesis, and motivating the new experimental scheme through a numerical study. By computing the thermal properties for different sizes of the Bose-Einstein thermal baths (characterized by $N_e$), we explore a range of effective accelerations as defined by our theoretical setup in Section \ref{sec:ts}. To achieve this, we numerically compute the eigenspectrum of the Hamiltonian matrix presented in Eq. \eqref{eq:Hamiltonian} for sixteen different selected values of $N_e$. This allows us to build the partition function for the sixteen systems. The partition function $Z$ holds probabilistic information about all possible configurations that the system can possess. For a canonical system in thermal equilibrium, the partition function is given by \cite{Gibbs_2010}:
\begin{equation}
Z(\beta) = \sum_{l=1}^{N_e+1} e^{-\beta\varepsilon_l} ,
\end{equation}
where $\beta = (k_B T)^{-1}$ is the Boltzmann factor, and $\{\varepsilon_l\}_{l=1...N_e+1}$ are the eigenvalues of the $(N_e+1)\times(N_e+1)$ Hamiltonian matrix in Eq. \eqref{eq:Hamiltonian}, which are computed numerically (see publicly available code in \cite{Sutemp2025}). To compute the critical temperature for each of the sixteen systems, we first need to estimate the heat capacity $C$. We do this by calculating the internal energy \cite{Gibbs_2010,pathria2021statistical}:
\begin{equation}
\overline{E}(\beta) = \text{Tr}(\rho H) = Z^{-1}(\beta) \sum_{l=1}^{N_e+1} e^{-\beta\varepsilon_l} \langle\psi_l|H |\psi_l\rangle .
\end{equation}
where $\{|\psi_{l}\rangle\}_{l=1...N_e+1}$ are the eigenstates of the $(N_e+1)\times(N_e+1)$ Hamiltonian matrix in Eq. \eqref{eq:Hamiltonian}. Then, we find the heat capacity by taking the derivative of the internal energy with respect to temperature \cite{Gibbs_2010,pathria2021statistical}:
\begin{equation}
C(\beta) = \frac{\partial\overline{E}(\beta)}{\partial T} = k_B\beta^2 \frac{\partial^2 \ln Z(\beta)}{\partial \beta^2} .
\end{equation}
\begin{figure}[t]
    \centering
    \includegraphics[width=\linewidth]{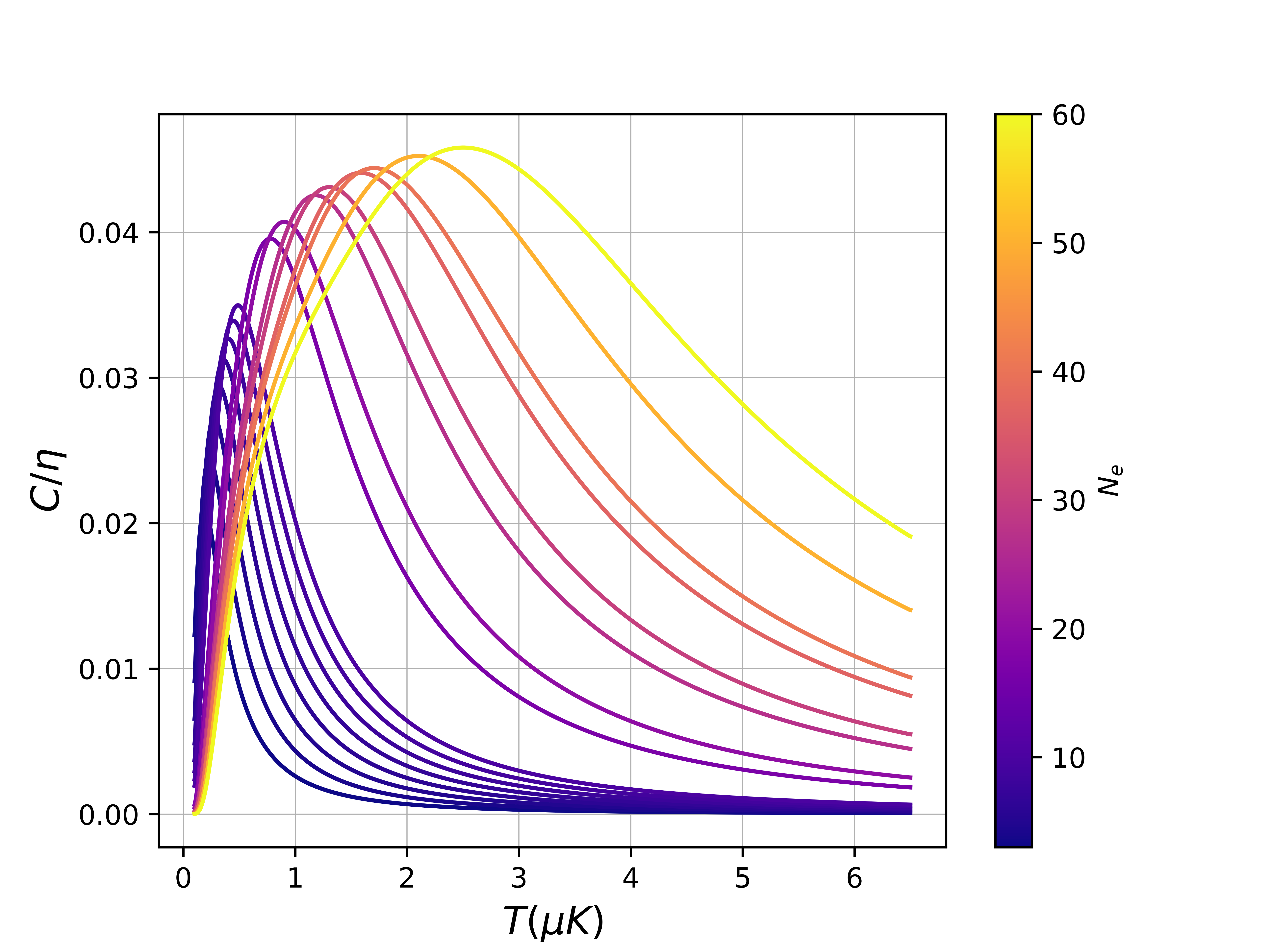}
    \caption{ Numerical estimation of the normalized heat capacity $C$  for sixteen different Bose-Einstein thermal baths (characterized by $N_e$) as a function of temperature $T$, where  $\eta = \hbar g_{ch}$ and $g_{ch}$ is the characteristic coupling frequency given by Eq. \eqref{eq:gprime}. The characteristic acceleration is fixed at $A_{ch} \approx 1.07\times10^{14} m/s^2$, resulting in a characteristic temperature and $\eta \approx 10 , neV$. The plot shows that the heat capacity of each system increases with temperature, reaching a maximum at the critical temperature $T_c$, which increases with the maximum number of excited atoms $N_e$. For example,  for  $N_e=3$, the estimated critical temperature is $T_c\approx1.89\times10^{-7}$, where for the last snapshot with $N_e=60$ we got   $T_c\approx2.50\times10^{-6}$. This figure illustrates the method for numerically determining the critical temperature for each Bose-Einstein thermal bath from the heat capacity via computing the gradient of each graph.}
    \label{fig:heat capacity VS tempreature}
\end{figure}
\\
Figure \ref{fig:heat capacity VS tempreature} shows the numerical estimation of the heat capacity for sixteen Bose-Einstein thermal baths as a function of temperature $T$, where $\eta = \hbar g_{ch}$. To compare our results with the data in Ref. \cite{Hu2018QuantumSO}, the characteristic acceleration was fixed to $A_{ch} \approx 1.07\times10^{14} m/s^2$. Figure \ref{fig:heat capacity VS tempreature} shows that the heat capacity of each system increases with temperature, indicating effective heat absorption. The heat capacity reaches its maximum value at a critical temperature $T_c$, indicating the phase transition. Mathematically, $T_c$ is defined by the conditions \cite{pathria2021statistical}:
\begin{equation}
\frac{\partial C(T )}{\partial T }\Big|_{T=T_c} = 0 \quad \text{and} \quad \frac{\partial^2 C(T )}{\partial T ^2 }\Big|_{T=T_c} < 0 .
\end{equation}
For the sixteen systems considered, the critical temperature increases with the maximum number of excited atoms. According to our hypothesis ($T_U \equiv T_c$), the critical temperature of each bath should correspond to a specific Unruh temperature associated with a particular acceleration. To further validate this, we will now examine the average number of the phononic excitations at these critical temperatures, as our model predicts a relationship between $\bar{n}(T_c)$ and the acceleration. The average number of the phononic excitations for each snapshot $s$, $\bar{n}_{s}(\beta)$, is given by \cite{Gibbs_2010}:
\begin{equation}
\label{eq:anpex}
\overline{n}_s(\beta)= \text{Tr}(\rho\hat{N}) = Z^{-1}(\beta) \sum_{l=1}^{N_e+1} e^{-\beta\varepsilon_l} \langle\psi_l| \hat{N} |\psi_l\rangle ,
\end{equation}
where $\hat{N}$ is the number operator. The next step is to use Eq. \eqref{eq:anpex} to numerically compute the average number of the phononic excitations at the critical temperature $\bar{n}_{s}(T_c)$  for the sixteen thermal baths and correlate these values with their corresponding critical temperature $T_c$. By plotting the set of scattered points $(\bar{n}(T_c), T_c)$, we can simulate the Unruh temperature through the proportionality between the critical temperature of each Bose-Einstein thermal bath and the average number of the phononic excitations at the critical temperature.
\begin{figure}[t]
    \centering
    \includegraphics[width=\linewidth]{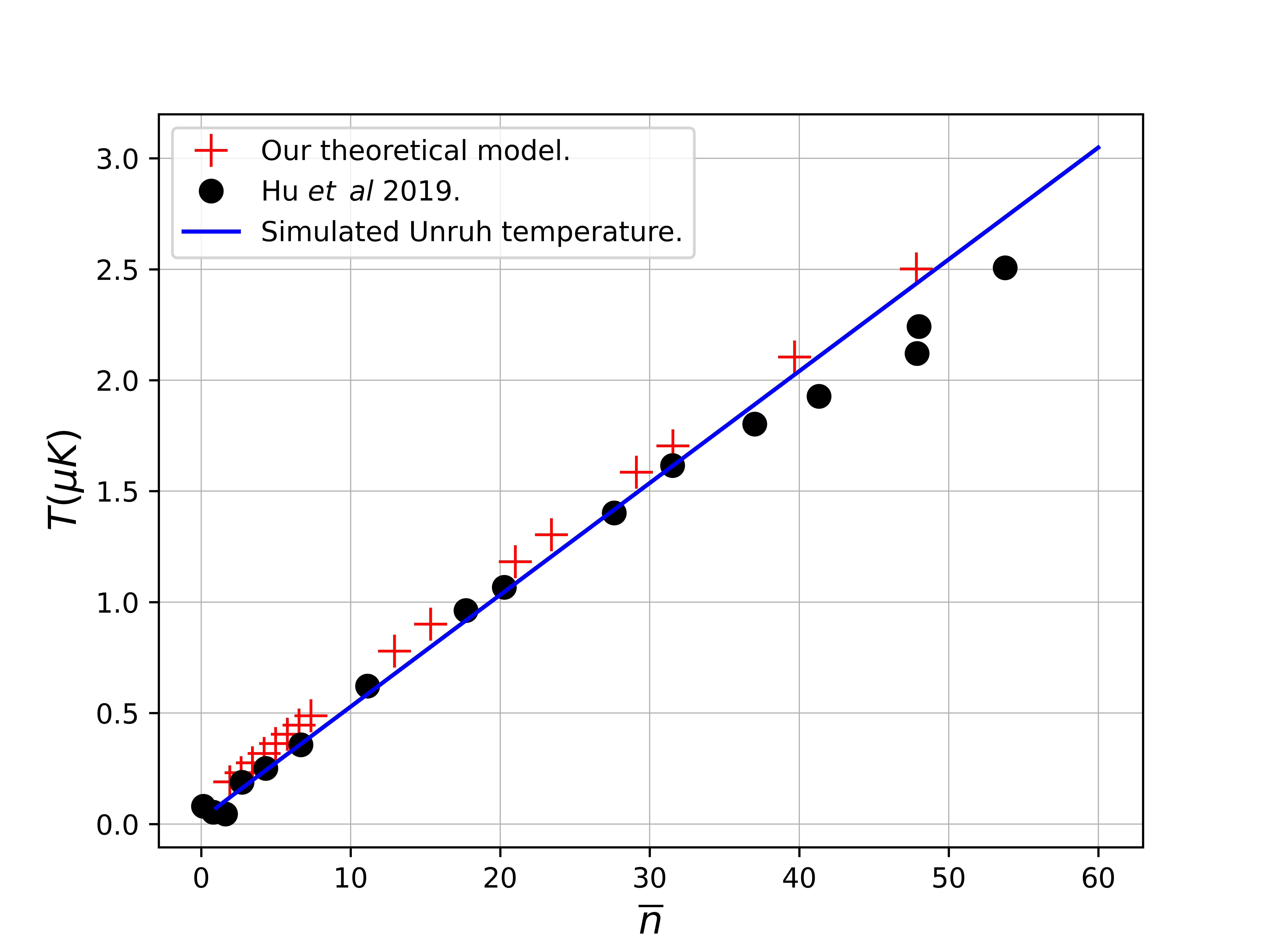}
     \caption{Simulation of the Unruh temperature through the proportionality between the critical temperature ($T_c$) of each Bose-Einstein thermal bath and the average number of the phononic excitations at the critical temperature ($\bar{n}_{s}(T_c)$). The scattered red crosses represent $(\bar{n}_{s}(T_c), T_c)$ for the sixteen thermal baths, representing our simulated Unruh temperature formula, yielding a ratio  of $\kappa\approx1.21,pK \cdot s$. This figure compares our simulation results to the experimental data from Ref. \cite{Hu2018QuantumSO} (black filled circles) corresponding to ($\bar{n}_{exp} \,,T_{exp}$), resulting a ratio of  $\kappa\approx 1.17(7)\, pK\cdot s$ \cite{Hu2018QuantumSO}, and to the theoretical Unruh temperature given by $T=\frac{\kappa }{c}A$ (blue solid line) where $A$ is the simulated Unruh acceleration which can be found as a function of the effective mean population $\bar{n}$ as shown in Eq. \eqref{eq:TME}, where it gives a ratio of $\kappa\approx 1.22\, pK\cdot s$. The agreement between the theoretical curve, the experimental data and the results of our simulations supports the hypothesis $T_U \equiv T_c$.}
    \label{fig:SUT}
\end{figure}
In Figure \ref{fig:SUT} we compare the results of our theoretical model simulating the experiment of Hu \textit{et al.} to the simulated Unruh temperature that takes the following form \cite{Hu2018QuantumSO}:
\begin{equation}
\label{eq:TME}
    T_{U} =\frac{\kappa}{c}A\quad ,\quad  \text{with \ } A =\frac{2\pi cE}{\hbar\ln(1+\frac{1}{\bar{n}})} ,
\end{equation}
where $A$ is the simulated effective acceleration, function of the effective mean population $\bar{n}$. In our model, $A$ is simulated through $\bar{n}_{s}(T_c)$, the average number of the phononic excitations for each snapshot at the critical temperature with $s=\overline{1,16}$. For large effective mean population $\bar{n} \gg 1$ the simulated Unruh acceleration scales linearly with $\bar{n}$ so $A=\frac{2\pi c E }{\hbar}\,\bar{n}$. Moreover, in Eq.\eqref{eq:TME}  $E=\hbar\omega/2$ is the effective kinetic energy of each atom, where the modulation frequency takes the experimental value used in Ref.   \cite{Hu2018QuantumSO}, $\omega=2\pi \times 2.1\, $kHz.
Our simulation results presented in Figure \ref{fig:SUT} indicate that our model reproduces the Unruh temperature formula, which takes the form $T=\kappa A/c $ where $\kappa=\hbar/2\pi k_B$. The results of our model predict a relationship between the critical temperature of these multiple Bose–Einstein thermal baths and the mean population $\bar{n}$. However, obtaining a closed-form analytical expression that relates the critical temperature to $\bar{n}$ is challenging, as no simple formula has yet been found to calculate the eigenvalues of the Hamiltonian that describe the systems used to simulate the new experimental setup, which take the tri-diagonal form with variable matrix elements. Where in \cite{trefethen2005spectra}, the authors clarify that while some structured matrices such as Toeplitz forms allow closed-form eigenvalue expressions, more general or variable-coefficient tridiagonal matrices may exhibit highly nontrivial spectral behavior and typically require numerical treatment for analysis. Moreover, a universal expression  for determining the critical temperature of discrete systems governed by the Hamiltonian matrix employed in our model is not yet found.
Furthermore, Figure \ref{fig:SUT} shows a comparison between our results and the experimental data from Ref. \cite{Hu2018QuantumSO}, where they obtained a value of $\kappa \approx 1.17(7) \, pK \cdot s$. Compared to the simulated Unruh temperature (blue solid line) which have a ratio  of   $\kappa \approx 1.22 \, pK \cdot s$, the fitting of our data points (red crosses) exhibits a ratio of $\kappa \approx 1.21 \, pK \cdot s$, providing direct support for our hypothesis. It is worth noting that our model inspire an experimental setup that can lead to a more accurate simulation without relaying on extensive  experimental resources.

\section{Conclusion}
In this work, we have successfully demonstrated a novel theoretical framework for simulating the Unruh temperature by establishing a direct link to the critical temperature of multiple Bose-Einstein thermal baths. By treating the evolution of a driven BEC as a sequence of quasi-static snapshots, each representing an independent thermal bath, we were able to derive the critical temperature based on the heat capacity, which was numerically estimated from the partition function. Our key finding is the significant agreement between the relationship derived from our model (involving the average number of the phononic 
excitations at the critical temperature and the acceleration) and the well-established Unruh temperature formula. This agreement strongly supports our hypothesis that the Unruh temperature can be effectively simulated through the critical temperature of these Bose-Einstein thermal baths.
This approach provides a new experimental scheme that holds the potential for a scalable cost-effective and time-efficient simulations---which would be necessarily quantum---compared to the significant experimental resources required for directly observing the Unruh effect through extreme accelerations or simulating it via highly sensitive and delicate quantum systems provided in other experimental setups. Moreover, our model further strengthens the perspective of analogue gravity, providing another theoretical lens through which to understand the deep connections between condensed matter physics and spacetime phenomena. The snapshot approach to analyzing the dynamics of the driven BEC offers a unique methodology for studying complex, evolving quantum systems by applying concepts from equilibrium statistical mechanics. 
This work opens avenues for future research exploring the limits and applicability of this analogy under different system parameters and for investigating other relativistic phenomena beyond the Unruh effect. Furthermore, it will encourage the development of experimental schemes, either through digital quantum simulations or Bose-Einstein condensate (BEC) experiments.

%%%%%%
\section*{Acknowledgments}
This document has been produced with the financial assistance of the European Union (Grant no. DCI-PANAF/2020/420-028), through the African Research Initiative for Scientific Excellence (ARISE), pilot program. ARISE is implemented by the African Academy of Sciences with support from the European Commission and the African Union Commission. The contents of this document are the sole responsibility of the author(s) and can under no circumstances be regarded as reflecting the position of the European Union, the African Academy of Sciences, and the African Union Commission.
We are grateful to the Algerian Ministry of Higher Education and Scientific Research and DGRST for the financial support.We thank Mohamed Messaoud Louamri for helpful discussion on numerical simulation included in this study.
\bibliographystyle{plain}
\bibliography{Ref.bib}

\section*{Invariance of the Hamiltonian Under the Simulated Rindler Transformation}
\label{sec:Appendix A}
In this section, we will demonstrate the invariance of the Hamiltonian presented in Eq. \eqref{eq:second-quatized-hamiltonian}  under the transformation described in Ref. \cite{Hu2018QuantumSO} , which can be understood as a Rindler coordinate transformation, i.e., a transformation to a boosted frame where the annihilation and creation operators become $a_k(\tau)$ and $a^\dagger_{-k}(\tau)$ respectively. We begin by defining the Rindler operator \cite{Hu2018QuantumSO}:
\begin{equation}
\label{eq:rindler}
    \hat{\mathcal{R}}=e^{g\tau\sigma_{x}}.
\end{equation}
The annihilation and creation operators for an atom with momentum $k$ transform as:
\begin{align*}
\begin{bmatrix}
   a_{k}(\tau) \\
    a_{-k}^{\dagger}(\tau) \\
\end{bmatrix}&=e^{g\tau\sigma_{x}}\begin{bmatrix}
   a_{k}(0) \\
    a_{-k}^{\dagger}(0) \\
\end{bmatrix}\\
    \begin{bmatrix}
   a_{k}(\tau) \\
    a_{-k}^{\dagger}(\tau) \\
\end{bmatrix}&=\begin{bmatrix}
    \cosh(g\tau) & \sinh(g\tau) \\
    \sinh(g\tau) & \cosh(g\tau) \\
\end{bmatrix}
\begin{bmatrix}
   a_{k}(0) \\
    a_{-k}^{\dagger}(0) \\
\end{bmatrix}\\
    \begin{bmatrix}
   a_{k}(\tau) \\
    a_{-k}^{\dagger}(\tau) \\
\end{bmatrix}&=\begin{bmatrix}
   \cosh(g\tau)a_{k}(0)+\sinh(g\tau)a_{-k}^{\dagger}(0) \\
    \sinh(g\tau)a_{k}(0)+\cosh(g\tau)a_{-k}^{\dagger}(0) \\
\end{bmatrix}.
\end{align*}

Also, we have:
\begin{align*}
    \begin{bmatrix} a_{k}^{\dagger}(\tau) & a_{-k}(\tau)  \end{bmatrix}&=\begin{bmatrix} a_{k}^{\dagger}(0) & a_{-k}(0)  \end{bmatrix}\begin{bmatrix}
    \cosh(g\tau) & \sinh(g\tau) \\
    \sinh(g\tau) & \cosh(g\tau) \\
    \end{bmatrix}\\
    \begin{bmatrix} a_{k}^{\dagger}(\tau) & a_{-k}(\tau)  \end{bmatrix}&=
    [ \cosh(g\tau)a_{k}^{\dagger}(0)+\sinh(g\tau)a_{-k}(0) \\&\quad\quad\quad\,\sinh(g\tau)a_{k}^{\dagger}(0)+\cosh(g\tau)a_{-k}(0)].
\end{align*}
We can use the above equations to get:
\begin{align*}
    a_{k}^{\dagger}(\tau)a_{-k}^{\dagger}(\tau)&=\cosh(g\tau)\sinh(g\tau)a_{k}^{\dagger}(0)a_{k}(0)\\%$+\cosh^{2}(g\tau)a_{k}^{\dagger}(0)a_{-k}^{\dagger}(0)\\
    &+
    sinh^{2}(g\tau)a_{-k}(0)a_{k}(0)\\&+\sinh(g\tau)\cosh(g\tau)a_{-k}(0)a_{-k}^{\dagger}(0)\\
    a_{k}(\tau)a_{-k}(\tau)&=\cosh(g\tau)\sinh(g\tau)a_{k}(0)a_{k}^{\dagger}(0)\\&+\cosh^{2}(g\tau)a_{k}(0)a_{-k}(0)\\
    &+\sinh^{2}(g\tau)a_{-k}^{\dagger}(0)a_{k}^{\dagger}(0)\\&+\sinh(g\tau)\cosh(g\tau)a_{-k}^{\dagger}(0)a_{-k}(0)
\end{align*}
Now, by fixing $\comm{a_{k}(0)}{a^{\dagger}_{k^{'}}(0)}=\delta_{kk^{'}}$ and using the identity $\cosh^{2}(x)-\sinh^{2}(x)=1$, we can finally deduce that:
\begin{equation}
    a_{k}^{\dagger}(\tau)a_{-k}^{\dagger}(\tau)-a_{k}(\tau)a_{-k}(\tau)=a_{k}^{\dagger}(0)a_{-k}^{\dagger}(0)-a_{k}(0)a_{-k}(0)
\end{equation}

Thus, the interaction Hamiltonian we used is invariant under the Rindler transformation generated by the operator in \eqref{eq:rindler}, in addition we can check the invariance of  the commutation relation :
\begin{equation}
    \comm{a_{k}(\tau)}{a^{\dagger}_{k^{'}}(\tau)}=\comm{a_{k}(0}{a^{\dagger}_{k^{'}}(0)}=\delta_{kk^{'}}\ .
\end{equation}
\end{document}